\documentstyle[12pt]{article}
\catcode`\@=11

\catcode`\@=11

\def\section{\@startsection{section}{1}{\z@}{3.5ex plus 1ex minus
   .2ex}{2.3ex plus .2ex}{\large\bf}}
\def\thesection{\arabic{section}.}

\def\appendix{\setcounter{section}{0}
        \def\thesection{Appendix.}
        \def\theequation{\Alph{section}.\arabic{equation}}}

\long\def\@makefntext#1{\parindent 0cm\noindent
\hbox to 1em{\hss$^{\@thefnmark}$}#1}

\def\eqnarray{\let\@currentlabel=\theequation\refstepcounter{equation}
    \global\@eqnswtrue
    \global\@eqcnt\z@\tabskip\@centering\let\\=\@eqncr
    $$\halign to \displaywidth\bgroup\@eqnsel\hskip\@centering
      $\displaystyle\tabskip\z@{##}$&\global\@eqcnt\@ne
       \hfil${{}##{}}$\hfil
      &\global\@eqcnt\tw@ $\displaystyle\tabskip\z@{##}$\hfil
       \tabskip\@centering&\llap{##}\tabskip\z@\cr}
\def\lefteqn#1{\hbox to 4\arraycolsep{$\displaystyle #1$\hss}}
\def\rref#1{(\ref{#1})}
\newcommand{\beq}{\begin{equation}}
\newcommand{\eeq}{\end{equation}}

\begin{document}

\begin{titlepage}
\vspace{.5in}
\begin{flushright}
UCD-98-14\\
astro-phy/9808021\\
August 1998\\
revised December 1998\\
\end{flushright}
\vspace{.5in}
\begin{center}
{\Large\bf
Remarks on the ``New Redshift Interpretation''}\\
\vspace{.4in}
{S.~C{\sc arlip}\footnote{\it email: carlip@dirac.ucdavis.edu}\\
       {\small\it Department of Physics}\\
       {\small\it University of California}\\
       {\small\it Davis, CA 95616}\\{\small\it USA}}\\
\vspace{1ex}
{\small and}\\
\vspace{1ex}
{R.~S{\sc cranton}\footnote{\it email: scranton@oddjob.uchicago.edu}\\
        {\small\it Department of Astronomy and Astrophysics}\\
        {\small\it University of Chicago}\\
        {\small\it Chicago, IL 60637}\\
        {\small\it USA}}\\
\end{center}

\vspace{.5in}
\begin{center}
\begin{minipage}{4.5in}
\begin{center}
{\large\bf Abstract}
\end{center}
{\small In a recent article in {\em Modern Physics Letters} {\bf A},
Gentry proposed a new static cosmological model that seeks to explain
the Hubble relation as a combination of gravitational and Doppler red
shifts.  We show that Gentry's model, although supposedly based on
general relativity, is inconsistent with the Einstein field equations;
that it requires delicate fine tuning of initial conditions; that it
is highly unstable, both gravitationally and thermodynamically; and
that its predictions disagree clearly with observation.
}
\end{minipage}
\end{center}
\end{titlepage}
\addtocounter{footnote}{-2}

In a recent article in {\em Modern Physics Letters} {\bf A}, Gentry
has proposed a new static cosmological model, the ``New Redshift
Interpretation'' (NRI), which seeks to explain the Hubble relation as
a combination of gravitational and Doppler red shifts \cite{Gentry}.
In this paper, we provide critique of the proposal, which, as we shall
demonstrate, is inconsistent with both theory and observation.  We
shall not attempt a complete review, and will occasionally use
order-of-magnitude estimates rather than more precise computations,
since these are already sufficient to demonstrate that the model
fails.

We begin with a very brief summary of Gentry's proposal.  The model is
described as being ``based on a universe governed by static-space-time
general relativity,'' and cannot use the expansion of spacetime to
explain the Hubble red shift-distance relationship.  Instead, it
interprets cosmological red shifts as a combination of gravitational
red shifts and ordinary Doppler shifts coming from motion in a static
geometry.  While the spatial geometry is never explicitly given, a
number of expressions---for example, the integration measure in equation
(3) of the reference---indicate that Gentry is considering a Euclidean,
or at least nearly Euclidean, spatial metric.

To be consistent with observed isotropy of matter and cosmic microwave
background radiation, Gentry's model is geocentric, placing our galaxy
near the center of a static spherical ball of matter with a radius
$R\sim 1.4\times 10^{10}$ light years.  This matter consists of two
components: ordinary pressureless matter---galaxies---with a density
$\rho_m$, and vacuum energy (that is, a cosmological constant) with a
density $\rho_v$ and a pressure $p_v = -\rho_v$.  The gravitational
potential varies with distance from the center, and the resulting
gravitational red shift thus depends on distance from the Earth.

Gentry derives a ``Hubble constant'' $H$  of
\beq
H^2 = {4\pi G\over3}(2\rho_v-\rho_m) .
\label{a1}
\eeq
He computes the red shift-distance relation, combining gravitational and
Doppler red shifts, to be
\beq
z = {{1 + Hr}\over\sqrt{1 - (2+u_g^2)H^2r^2}} - 1
\label{a2}
\eeq
where $u_g$ is a ratio of transverse to radial velocities.  (We have
used units $c=1$.)

To explain cosmic microwave background radiation, Gentry supposes that
his ball of matter is surrounded by a thin shell of hydrogen at a
temperature of $5400\ {\mathrm K}$.  The resulting black body radiation
must be gravitationally red shifted to the observed $2.7\ {\mathrm K}$
at the center of the ball.  Using a matter density of $\rho_m\sim
2\times 10^{-31}\ {\mathrm g/cm}^3$ and adjusting the vacuum energy
to give the required CMBR red shift, he obtains a value of $\rho_v\sim
8.8\times 10^{-30}\ {\mathrm g/cm}^3$.

The characteristic time scale in the NRI model is given by the radius
$R$, which satisfies $HR\approx1$.  (From equation (1) of \cite{Gentry},
it may be seen that $H^2R^2 = 1 - 2.5\times10^{-5}$.)  Even if we ignore
time for equilibriation, the universe must have an age of at least $H^{-1}$
to fill up with radiation from the hydrogen shell.

\section{General Relativity}

We first note that Gentry's model, although supposedly based on general
relativity, is in fact inconsistent with the Einstein field equations.
The matter in the model is a perfect fluid, that is, it has no shear
stresses.  Any static, spherically symmetric perfect fluid in general
relativity is governed by a mass relationship,
\beq
{dm(r)\over dr} = 4\pi r^2 \rho
\label{b1}
\eeq
and the Oppenheimer-Volkov equation,
\beq
{dp\over dr} = - {(\rho + p)(m(r) + 4\pi r^3 p)\over r(r-2m(r))} ,
\label{b2}
\eeq
where $\rho$ and $p$ are the combined density and pressure of all of
the components of the fluid \cite{MTW,Weinberg}.  For Gentry's model,
$\rho = \rho_m + \rho_v$ and $p = -\rho_v$, and the constancy of $p$
implies from \rref{b2} that either $\rho = -p$ or $m(r) = -4\pi r^3p$.
In the latter case, we can differentiate again and use \rref{b1},
obtaining
\beq
\rho_m = \left\{ \begin{array}{ll} 0 & \mbox{for $\rho = -p$}\\
                             2\rho_v & \mbox{for $m(r) = -4\pi r^3p$} .
         \end{array}\right.
\label{b3}
\eeq
The first case gives an matter-free universe with a cosmological constant,
and is, in fact, de Sitter space.  The second yields a universe in which
Gentry's ``Hubble constant'' \rref{a1} vanishes; it may be recognized
as the Einstein static universe.  Neither is consistent with Gentry's
desired values of $\rho_m$ and $\rho_v$.  Conversely, the static NRI
model is not consistent with general relativity.

One might try to save the model by allowing radial dependence of $\rho$.
Such density variations are limited by observations of the interaction
of cosmic microwave background radiation with distant matter, which
indicate that the universe is radially homogeneous to within about 10\%
at the horizon scale \cite{Goodman,Holzapfel}.  But even ignoring these
limits, $r$-dependence of $\rho$ will not help as long as the pressure
$p$ comes from a cosmological constant.  Indeed, ``vacuum energy'' is
always constrained by the Bianchi identities to be constant, and we
only needed constancy of $p$, and not $\rho$, to obtain \rref{b3}.
The conclusion therefore stands.

Alternatively, one might drop the requirement of time-independence,
although this would largely defeat the purposes of the model.\footnote{%
Gentry explicitly requires only time-independence of the metric, and not 
of the matter density and pressure.  But a static metric in the Einstein 
field equations necessarily implies a static stress-energy tensor.  
Conversely, it is a trivial consequence of the field equations that a 
time-dependent matter distribution automatically implies a time-dependent 
geometry, so the claim in reference \cite{Gentry2} that the model ``is 
definitely not a static description'' but is nevertheless ``governed by 
static-space-time general relativity'' is inconsistent.}  The relevant 
Einstein tensor is then given in exercise 14.16 of reference \cite{MTW}.  
If $\rho$ and $p$ are taken to be independent of $r$, it may be shown that 
the resulting field equations reduce to those for a standard Friedman-%
Lema{\^\i}tre-Robertson-Walker metric.  This is essentially because the 
radial constancy of $p$ and $\rho$ in the region inside Gentry's hydrogen 
shell together with spherical symmetry imply spatial homogeneity in this 
region, and isotropy around the origin plus spatial homogeneity are sufficient 
to determine that the metric is locally of the FLRW form \cite{Weinberg}.
The model thus reduces to a spherical piece cut out of a standard
FLRW big bang model, surrounded by a now-superfluous shell of hydrogen.
Such solutions have been described by Smoller and Temple \cite{Smoller}.
They most definitely do not describe static spacetimes; the metric in the 
interior is indistinguishable from a standard FLRW metric.  This observation
explains equation \rref{b3}: the de Sitter and Einstein static universes
are the only FLRW cosmologies with constant pressure, the latter because
it is genuinely static and the former because the energy of the vacuum
remains proportional to the volume.

\section{Fine Tuning}

Let us ignore these theoretical problems for the moment, and assume
that the model of reference \cite{Gentry} is a solution of some as yet
unknown theory that replaces general relativity.  This will cause
some problems, since it is difficult to make predictions from the
model without a clear theoretical framework, but we will proceed as
far as we can.  Our next observation is that the NRI model requires
fine tuning of parameters to reproduce anything close to the observed
universe.  This is first evident in the choice of temperature of the
hydrogen shell, which is apparently completely arbitrary and must be
adjusted to reproduce the observed microwave background temperature.
In the standard big bang model, by contrast, the cosmic microwave
background temperature is strongly constrained by primordial
nucleosynthesis \cite{Peebles}.  Since Gentry's model contains no
mechanism for primodial nucleosynthesis, no such constraint exists.

Further careful tuning is required to obtain the velocity-distance
relationship described after equation (4) of \cite{Gentry}.  Gentry
argues that a galaxy at distance $r$ from the center of the universe
experiences an acceleration
\beq
{\ddot {\bf r}} = -{GM(r)\over r^2}{\hat{\bf r}}  \quad\hbox{with}\quad
M(r) = {4\pi r^3\over3}(\rho_m - 2\rho_v) ,
\label{c1}
\eeq
where $\rho_m - 2\rho_v = \rho + 3p$ is the effective density that
contributes to the gravitational interaction in general relativity.
In a vacuum energy-dominated spacetime, $M(r)$ is negative, and the
resulting acceleration is radially outward.  Gentry states that equation
\rref{c1} implies that $r = r_0\exp\{Ht\}$, where $H$ is given by
\rref{a1}, and thus ${\dot r} = Hr$.

In fact, the general solution of \rref{c1} is
\beq
{\bf r}
  = {1\over2}\left[ \left({\bf r}_0 + {{\bf v}_0\over H}\right)e^{Ht}
  + \left({\bf r}_0 - {{\bf v}_0\over H}\right)e^{-Ht} \right] ,
\label{c2}
\eeq
where ${\bf r}_0$ and ${\bf v}_0$ are the position and velocity at $t=0$.
The solution of reference \cite{Gentry} is recovered only if we require
that ${\bf v}_0 = H{\bf r}_0$.  That is, the Hubble velocity-distance
relationship is obtained only if it is imposed as an initial condition.
Observe that this is a distinct initial condition for each galaxy: we
must separately require that ${\bf v}_0 = H{\bf r}_0$ for each object
in the universe.  Note also that this fine tuning is absolutely
essential for the rest of the argument of reference \cite{Gentry}, since
without the resulting Doppler shift, equation (3) of that reference would
give red shifts $z\propto r^2$ rather than $z\propto r$ for small $r$.

\section{Stability and Consistency}

We next examine several issues of stability and self-consistency of the
NRI model, one involving the interior matter and others concerning the
exterior hydrogen shell.  We start with the observation that the ``Hubble
flow'' \rref{c2}, with Gentry's initial conditions ${\bf v}_0 = H{\bf r}_0$,
is not consistent with a static, radially constant matter density $\rho_m$.
If we assume that $\rho_m$ is independent of $r$---as required, for
instance, for the force computation described in equation \rref{c1}---and
set ${\bf v} = H{\bf r}$, the continuity equation gives
\beq
{\partial\rho_m\over\partial t} = -\nabla\cdot(\rho_m{\bf v})
= -3H\rho_m
\label{d1}
\eeq
i.e.,
\beq
\rho_m(t) = \rho_m(0)e^{-3Ht} .
\label{d2}
\eeq
The matter density thus drops by a factor of $20$ in the time $H^{-1}$ it
takes for the NRI universe to fill up with microwave background radiation.

Of course, equation \rref{d2} is inconsistent with the assumption of time
independence used to solve \rref{c1}; to do the computation correctly, one
must treat \rref{c1} and \rref{d1} as coupled equations.  It is not hard
to show that the result asymptotically approaches \rref{d2}, with $H$
obtained from the definition \rref{a1} with $\rho_m=0$.  One can obtain a
long period of slowly varying density by adjusting the initial value of
$\rho_m$ to be nearly $2\rho_v$, but the effective ``Hubble constant'' is
then very small.  In a sense, this is merely an awkward Newtonian statement
of the well-known instability of the Einstein static universe: density
is constant if $\rho_m=2\rho_v$, but any small perturbation leads to an
asymptotically empty de Sitter space.

The only way we can see to avoid this time dependence would be to drop
the continuity equation \rref{d1} and the consequent conservation of
matter.  This would require a return to something like the largely
discredited steady state cosmology of Bondi, Gold, and Hoyle (see section
14.8 of reference \cite{Weinberg}).  As we shall see below, however,
the NRI model is not the same as the old steady state model, but makes
substantially less accurate predictions for red shifts.

We next explore the question of the stability and the radiative properties
of the outer hydrogen shell in the NRI model.  Local gravitational stability
is not a problem: the thickness $R_S$ of the shell can be made less than
the Jeans length.  But the total mass of the shell is greater than the
(negative) mass it encloses.  Indeed, according to reference \cite{Gentry},
the shell mass is $M_S = 2\pi R^3(2\rho_v-\rho_m)$, so
\beq
M_{\hbox{\scriptsize\em tot}}
= M_S + M_{\hbox{\scriptsize\em interior}}
= {2\pi(2\rho_v - \rho_m)R^3\over3} > 0 .
\label{d5}
\eeq
Since the interior pressure due to the vacuum energy density is negative,
it will not stabilize the shell.  The time scale for collapse of an
initially static shell goes as
\beq
t_{\hbox{\scriptsize\em collapse}}
\sim \sqrt{R\left(-{d^2R\over dt^2}\right)^{-1}} = {\sqrt{2}\over H} ,
\label{d6}
\eeq
where we have used equation \rref{c1} with $M$ replaced by
$M_{\hbox{\scriptsize\em tot}}$.  This is the same as the time scale
\rref{d2} for expansion of the interior; both indicate that the NRI
``Hubble constant'' is actually a scale for instability.

The estimate
\rref{d6} has ignored internal pressures in the shell, of course, but
it seems unlikely that such pressures can stabilize a thin ($R_S\ll R$)
shell.  In particular, while we have argued above that the NRI model
cannot be considered general relativistic, it is interesting to note
that equations \rref{a1} and \rref{d5} imply that the radius $R$ at which
the shell is located is almost exactly the Schwarzschild radius $R_0 = 2G
M_{\hbox{\scriptsize\em tot}}$ of the mass $M_{\hbox{\scriptsize\em tot}}$:
\beq
R = R_0(HR)^{-2} \approx (1 + 2.5\times10^{-5})R_0 .
\label{d7}
\eeq

We should also worry about the radiative properties of the hydrogen
shell.  To be opaque enough to radiate as a black body, the shell's
optical depth $\tau$ must satisfy
\beq
\tau = n\sigma R_S \gg 1 ,
\label{d3}
\eeq
where $n$ is the number density, $R_S$ is the shell thickness, and
$\sigma$ is the relevant hydrogen scattering cross-section.  In the
likely temperature regimes for the shell in the NRI model, $\sigma$ will
be the Thomson cross-section.  The mass of the shell must thus satisfy
\beq
M_S = {4\pi\over3}\left((R+R_S)^3 - R^3\right)nm_H \approx 4\pi R^2 R_S nm_H
\gg 4\pi R^2 m_H/\sigma ,
\label{d4}
\eeq
where $m_H$ is the mass of a hydrogen atom.  For Gentry's value of $R$,
this gives a requirement that $M_S\gg 3\times10^{24} M_\odot$.

{}From the discussion before equation (1) of reference \cite{Gentry}, on
the other hand, the hydrogen shell in the NRI model has a mass of only
$M_S\approx 1.3\times10^{23}M_\odot$, and thus will not act as a black
body.  Equivalently, substituting Gentry's value of $M_S$ into \rref{d4}
and \rref{d3}, we obtain an optical depth of only $\tau\approx.04$.  It
seems impossible to adjust parameters in the model to move this number
significantly upward without obtaining an unacceptably large value for
the Hubble constant $H$.

We must additionally consider radiative cooling of the hydrogen shell.
The interior of the shell may be in thermal equilibrium, but according
to reference \cite{Gentry}, the exterior spacetime is assumed to be nearly
vacuum.  By the Stefan-Boltzmann equation, we expect thermal energy to be
radiated at a rate
\beq
{dE_{\hbox{\scriptsize\em therm}}\over dt}
  = 4\pi R^2\,\sigma_{\hbox{\scriptsize\em SB}} T^4 ,
\label{d4a}
\eeq
where $\sigma_{\hbox{\scriptsize\em SB}}$ is the Stefan-Boltzmann constant.
The total thermal energy, on the other hand, is approximately ${3\over2}kT$
per hydrogen atom, or
\beq
E_{\hbox{\scriptsize\em therm}} = {3\over2}\left( {M_S\over m_H}\right)kT .
\label{d4b}
\eeq
The characteristic cooling time is thus
\beq
t_{\hbox{\scriptsize\em cooling}} \sim E_{\hbox{\scriptsize\em therm}}
  \left( {dE_{\hbox{\scriptsize\em therm}}\over dt} \right)^{-1}
  = {3kM_S\over 8\pi m_HR^2\sigma_{\hbox{\scriptsize\em SB}}T^3} .
\label{d4c}
\eeq
For the parameters of reference \cite{Gentry}, this yields a time on the
order of a second; the system is drastically unstable.

This computation is nonrelativistic, of course, and does not give the
true energy radiated to infinity.  Indeed, if the shell is inside its
event horizon, no radiation will escape to infinity.  This is irrelevant
to the question of thermal stability, however.  Equation \rref{d4c}
gives the correct cooling time in a frame comoving with the hydrogen
shell.  Provided that back-scattering is small---which it surely will
be if the exterior spacetime is nearly empty---what matters is not what
happens to the outgoing radiation after it is emitted, but merely the
rate at which it leaves the shell.  Even inside an event horizon, where
the ``outgoing'' radiation converges to the singularity, the shell
necessarily collapses more quickly than the radiation; relative to the
shell, the radiation remains outgoing, and continues to cool the shell.

\section{Observation}

Finally, let us turn to the crucial question of whether the model of
reference \cite{Gentry} agrees with observation.  Gentry argues that the
red shift-distance relation \rref{a2} reduces to the Hubble relation
$z=Hr$ ``for small $r$.''  This is true if ``small $r$'' is small
enough, but in fact, equation \rref{a2} leads to significant deviations
from linearity for rather small red shifts, in regimes in which such
deviations would be (and are not) seen.  This effect is minimized if
$u_g=0$, but even in that case, there is about a $10\%$ deviation
from linearity for $z=.1$, and a $50\%$ deviation for $z=.5$.

In fact, for $z<1$, the red shift-distance relation \rref{a2} with
$u_g=0$ is well approximated by
\beq
z \approx (1+z)Hr .
\label{e1}
\eeq
In Gentry's (apparently) Euclidean setting, an object with absolute
luminosity $L$ at radius $r$ will have an apparent luminosity of
\beq
\ell = {L\over4\pi r^2(1+z)} .
\label{e1a}
\eeq
The factor of $1+z$ in the denominator reflects photon energy loss due
to red shift; in an expanding spacetime, a second factor of $1+z$
would appear in the denominator, reflecting the diminished photon arrival
rate due to the stretching of the path length during travel, but this
factor will not occur in a nonexpanding spacetime model like Gentry's
\cite{Sandageb}.  The luminosity distance is thus
\beq
Hd_L(z) = Hr\sqrt{1+z} \approx {z\over\sqrt{1+z}} ,
\label{e1b}
\eeq
and the deceleration parameter $q_0$, given by \cite{Weinberg}
\beq
Hd_L = z + {1\over2}(1-q_0)z^2 + \dots ,
\label{e2}
\eeq
is $q_0=2$.  By way of reference, for an FLRW universe with $\Lambda=0$,
a deceleration parameter $q_0=2$ corresponds to a density four times
the critical value necessary for recollapse.   Even before the recent
discovery that Type Ia supernovae can be used as standard candles,
such a prediction was excluded by observation \cite{Sandageb,Sandage}.
It is now strongly ruled out by the supernova observations of Permutter
et al.\ \cite{Perl} and Garnavich et al.\ \cite{Garn}.  Indeed, equation
\rref{e1b} implies a bolometric magnitude of
\beq
m \approx 5\log_{10}z - 2.5\log_{10}(1+z) + \hbox{\em const.}\
  \approx 5\log_{10}z - 1.086z + \hbox{\em const.},
\label{e2a}
\eeq
and a cursory look at the Hubble diagrams of references \cite{Perl}
and \cite{Garn} shows that this prediction is in conflict with the data.

A further observational test comes from predicted number counts.  Gentry
argues that his model may explain the paucity of quasars at $z>4$,
essentially because nonlinearities in the red shift-distance relation
\rref{a2} imply that a shell of width $\Delta z$ at large $z$ contains
very little physical volume.  Unfortunately, though, this fall-off
actually becomes important at red shifts significantly smaller than
$z=4$.

In a static cosmological model, the number density $n$ of, say, quasars
can reasonably be expected to be constant.  The number of objects in a
shell of width $dr$ and a solid angle $d\Omega$ is thus $dN = 4\pi n
r^2drd\Omega$, where we have again assumed Euclidean spatial geometry.
The number of objects between red shifts $z$ and $z+dz$ thus satisfies
\beq
H^3 dN = 4\pi n (Hr)^2\, {d(Hr)\over dz} dzd\Omega .
\label{e3}
\eeq
For Gentry's NRI model, $Hr_{\hbox{\scriptsize\em NRI}}$ is given by
equation \rref{a2}, while for a standard spatially flat FLRW model, we
have \cite{Sandageb}
\beq
Hr_{\hbox{\scriptsize\em FLRW}} = 2\left( 1 - {1\over\sqrt{1+z}}\right) .
\label{e4}
\eeq
A straightforward computation then gives a ratio of number counts
\beq
F(z)
 = {dN_{\hbox{\scriptsize\em NRI}}\over dN_{\hbox{\scriptsize\em FLRW}}}(z)
 = {w^2(1-2w^2)^{5/2}\over(1+2w)}
   {(1+z)^{5/2}\over4(\sqrt{1+z}-1)^2} ,
\label{e5}
\eeq
where we define $w=Hr_{\hbox{\scriptsize\em NRI}}$ and consider it to be
a function of $z$ determined implicitly by equation \rref{a2}.

For small red shifts, the ratio \rref{e5} is approximately $1-z$, and
the difference betwen the NRI model and a standard FLRW model is probably
not presently observable.  But $F$ begins to fall substantially for
larger $z$: $F(z=1)\approx.39$, and $F(z=2)\approx.18$.  Far from
explaining the observed distribution of quasars---a sharp {\em rise\/}
between $z=0$ and $z=2$, followed by a fall-off at high $z$
\cite{Rees}---Gentry's model thus predicts a substantial {\em decrease\/}
between $z=0$ and $z=2$.

Next, we would like to quickly mention a few other observational
issues.  First, as noted above, Gentry's model contains no mechanism for
primordial nucleosynthesis.  The standard big bang model successfully
predicts light element abundances, and no serious alternative for the
production of the observed quantities of helium and deuterium is known
\cite{nucleo}.  The NRI model must, apparently, attribute this success
to coincidence.  Second, the model has no explanation for the fairly
good agreement between a variety of astrophysical ages and the Hubble
time \cite{Chaboyer,Cowan,Oswalt}.  In standard cosmological models,
the age of the universe sets a natural time scale for old objects, but
if there is no big bang, we might reasonably expect to see objects much
older than $H^{-1}$.  Third, as a ``static spacetime'' cosmology, the
NRI model predicts a dependence of surface brightness on red shift
that is very different from that of a model involving true expansion
\cite{Sandageb}.  While the ``Tolman test'' of the surface brightness-red
shift relation is quite difficult, there is now some evidence against
static models \cite{Pahre}.

Finally, let us briefly address one other issue raised in references 
\cite{Gentry2} and \cite{Gentry3}, the problem of energy conservation in 
cosmological expansion.  Gentry notes, correctly, that the electromagnetic 
energy of the cosmic microwave background is not conserved during expansion:
in a volume expanding along with the universe, the radiation energy goes as
$(1+z)^{-1}$, and the red shift represents a genuine loss of photon energy.
But there is nothing particularly ``cosmological'' about this loss---a photon 
rising in a static gravitational potential experiences a similar energy loss.  
In the laboratory, there is nothing mysterious about this phenomenon, which 
simply reflects the need to include gravitational potential energy in one's 
accounting.  Indeed, energy conservation can be used to derive the red shift 
(see, for instance, section 7.2 of reference \cite{MTW}).

In the cosmological context this energy accounting is more difficult, given
the well-known problems in defining a local gravitational energy density in
general relativity.  But one can use, for example, the quasilocal energy
of Brown and York \cite{Brown} to investigate the total energy, including
gravitational potential energy, in a region of an expanding universe.  A
quick computation shows that for a spatially flat FLRW model, the total 
energy inside a sphere of constant proper radius $R$ remains constant during
cosmological expansion.  The extension to spatially curved universes is
currently under investigation.

\section{Conclusion}

The ``New Redshift Interpretation'' of reference \cite{Gentry} is a radical
reformulation of cosmology, which seeks to challenge the foundations of
the standard big bang model.  Such attempts are worthwhile; they are,
after all, an important way to test existing theories.

This attempt, however, clearly fails.  Despite its initial appeal to
general relativity, the NRI model is inconsistent with the Einstein field
equations, and its theoretical foundations are unclear.  If we ignore
these difficulties, we find that the model requires delicate fine tuning,
including a simultaneous specification of the initial velocity of each
galaxy in the universe.  The proposed mechanism for producing cosmic
microwave background radiation fails, since the outer hydrogen shell
is too thin to act as a black body, and hot and thin enough to cool in
seconds from outgoing radiation.  The model is unstable against both
expansion of the interior matter and collapse of the outer hydrogen shell.
It contains no mechanism for primordial nucleosynthesis, and cannot
explain the absence of objects much older than $H^{-1}$.  And finally,
the predicted red shifts and quasar number counts clearly disagree with
observation.

\vspace{1.5ex}
\begin{flushleft}
\large\bf Acknowledgements
\end{flushleft}

We would like to thank Sverker Johansson and Mark Kluge for useful
suggestions.  This work was supported in part by National Science
Foundation grant PHY-93-57203 and Department of Energy grant
DE-FG03-91ER40674.

\end{document}